\documentclass[12pt]{iopart}
\usepackage{graphicx}
\usepackage{cite}

\bibdata{methodsPaper}
\bibstyle{plain}
\begin{document}

\title{An Unsupervised Method for Quantifying the Behavior of Interacting Individuals}
\author{Ugne Klibaite, Gordon J. Berman, Jessica Cande, David L. Stern, Joshua W. Shaevitz}

\address{Department of Physics and Lewis-Sigler Institute for Integrative Genomics, Princeton University}
\address{Department of Biology, Emory University}
\address{Janelia Research Campus, HHMI}

\ead{shaevitz@princeton.edu}

\begin{abstract}
Social behaviors involving the interaction of multiple individuals are complex and frequently crucial for an animal's survival. These interactions, ranging across  sensory modalities, length scales, and time scales, are often subtle and difficult to quantify. Contextual effects on the frequency of behaviors become even more difficult to quantify when physical interaction between animals interferes with conventional data analysis, e.g. due to visual occlusion. We introduce a method for quantifying behavior in courting fruit flies that combines high-throughput video acquisition and tracking of individuals with recent unsupervised methods for capturing an animal's entire behavioral repertoire.  We find behavioral differences in paired and solitary flies of both sexes, identifying specific behaviors that are affected by social and spatial context. Our pipeline allows for a comprehensive description of the interaction between multiple individuals using unsupervised machine learning methods, and will be used to answer questions about the depth of complexity and variance in fruit fly courtship. 
\end{abstract}

\section{Introduction}

Social behaviors are exhibited by a wide variety of animals ranging from invertebrates to humans. These behaviors, both mundane and elaborate, can be crucial for an animal's success in its natural environment. Observations of social behaviors have revealed patterns that are common across distant species, indicating that certain patterns of action are evolutionarily favored and conserved \cite{sturtevant1915experiments}. For example, aggressive tendencies may be selected for in a variety of species because they afford aggressive individuals more access to resources and mating opportunities, while extravagant courtship sequences may have evolved because they signal a higher level of fitness. \cite{lorenz2002aggression,tinbergen1963aims}

There has been significant growth in the past decade in our ability to probe animal behavior with the goal of discovering and quantifying ever-smaller differences between phenotypes. Interest in genetic and neuronal control of behavior has accelerated the development of many recording and analysis techniques, and many groups have contributed to the efforts to track and identify individual and social behaviors in animals with the help of automation \cite{dankert2009automated,simon2010new,de2012computerized,kabra2013jaaba,schaefer2012surveillance,perez2014idtracker}. We can now design experiments where single or multiple individuals are recorded over hours or even days, and each motion of the individual can be captured and compared \cite{ohayon2013automated,dell2014automated}. This kind of data can be used to estimate behavioral variety in an individual, and to make statistical comparisons across individuals, experimental contexts, strains, and species, as well as interactions between these individuals \cite{gautrais2012deciphering}. While these methods have greatly increased our understanding of the behavior of individuals, unsupervised techniques to enumerate and monitor the behaviors of socially-interacting individuals, including rare events that only happen in specific social contexts, are lacking. 

One complex and oft-studied social interaction is innate male courtship in fruit flies. Courting males orient towards a female, tap her with their forelegs, and give chase. During courtship, the male deploys a variety of species-specific behaviors such as auditory cues like singing and abdomen drumming, as well as chemical cues such as those transmitted by licking and tapping. Eventually, if the female is receptive, copulation occurs \cite{greenspan2000courtship,bastock1955courtship,spieth1974courtship}. However, in spite of its innate nature, courtship is highly variable between individual males, and the underlying basis of this variability is largely unknown \cite{demir2005fruitless}. One potential source of this variation in male courtship results from male-female interactions. Computational techniques have been used to track and quantify certain stereotyped patterns in male courtship behavior, such as chasing, lunging, and song \cite{arthur2013multi,branson2009high}. However, in spite of decades of intense research into fly courtship, female courtship-specific behaviors, aside from that of slowing down in response to male song, remain largely undescribed \cite{coen2014dynamic}. Further, how interactions between males and females shape courtship behavior and contribute to variation in innate male courtship is completely unknown. To address these questions, we need more sophisticated methods to probe the behavior of socially-interacting animals. 

In order to understand how male-female interactions shape courtship behavior, we must be able to keep track of every facet of the behavior of each of two courting individuals across a large number of trials. Because these features can be difficult to score by eye, especially at the same time across multiple individuals, we must turn to automated methods to track and compare behaviors over time. Human observation is limited by factors such as definition bias and lapses in attention when observing many experiments over long periods of time. Supervised machine learning methods can be used to more accurately and fully observe complex behaviors for almost unlimited amounts of time but typically suffer from similar definition bias \cite{levitis2009behavioural}. In order to create a link between subtle behavioral phenotypes and sexual selection, we need methods that can detect and define every possible behavior and quantify its dependence on specific environmental contexts. We currently do not know which behaviors or parts of courtship interactions are important for the emergence of complex traits, and require methods that can search the space of all behavior and eliminate human biases in order to find them. 

Our group recently developed a method to catalog and quantify behavior in the form of a two-dimensional histogram of postural dynamics \cite{berman2014mapping}. This method is used to organize and visualize the postural movements of an animal in an unsupervised manner.  One complication that occurs when attempting to use this method with interacting animals is the mutual visual occlusion that occurs when animals touch and their bodies overlap in the field of view. For courtship experiments in particular, the male and female flies exhibit many behaviors where they touch, including licking, tapping, and mounting. Here, we describe how we overcame this problem with computer-vision software that carefully handles the tracking and segmentation of individuals over the course of a multi-fly to. Specifically, we introduce a method for tracking and segmentation that preserves identities and as much of the fly bodies as possible, even when imaging views are obstructed due to interaction. This method allows us to generate a behavioral map based on the behavior of both individuals. In the case of courtship, this provides a basis of comparison between male and female behavior. We demonstrate this method by finding subtle behavioral differences, in both males and females, between isolated and paired flies. We describe the emergence of behavioral programs based on the relative distance and orientation between two courting flies. Additionally, we examine the particulars of the context in which a behavior is performed, with the goal of constructing better models of courtship. 

\begin{figure}
\begin{centering}
\includegraphics[width=.75\linewidth]{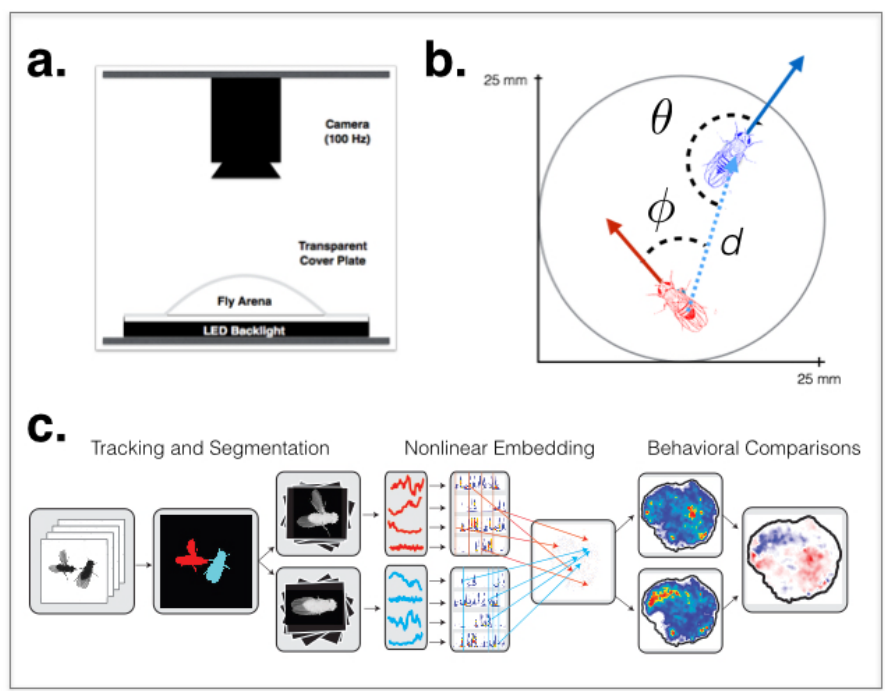}
\caption{ Experimental setup and analysis pipeline. a) Diagram of the recording setup used to take behavioral movies.  b) Schematic of angles and distances that capture the relative orientation of the male (red) and female (blue).  c) Pipeline of image analysis steps used to produce the joint behavioral distributions.}
\label{Figure 1}
\end{centering}
\end{figure}

\section{Behavioral Analysis of Courting Flies}
Our group  previously introduced an unsupervised method for discovering and cataloging the behaviors of individual animals using high-resolution movies \cite{berman2014mapping}. For courtship, the problem is more complex: we wish to study the joint behaviors of two interacting individuals recorded in a single movie, tracking both what the animals do and how their actions influence each other. The first step in analyzing interactions is to track and segment the two fly bodies from each other to create a movie for each individual (\S \ref{threshold}-\ref{movies}). This allows us to monitor the behavior of each animal independently in addition to other parameters such as their relative positions and orientations (Fig. \ref{Figure 1}b). These time series, which capture both behavior and context, can then be used to study behavioral interactions (\S \ref{maps},\ref{position}). 

\subsection{Calculating Image Thresholds}\label{threshold}

Due to a backlit imaging setup, our video recordings produce a silhouette image of the flies where light is completely occluded by the body and only partially occluded by the wings and limbs. We performed all image operations on inverse images (light body on dark background) so that the highest intensity pixels reside in the fly bodies, medium intensity pixels make up the wings and limbs, and the background is low intensity. We calculated a low threshold $T_F$, which is higher than most of the background noise, to locate large connected components in each image. A higher threshold, $T_B$, was calculated to discern pixels that belong to the body (thorax, head, and abdomen) as opposed to the appendages. One concern in segmenting movies with multiple flies is that small lighting anisotropies and filming irregularities may disrupt segmentation during frames where flies touch. In order to successfully segment multiple flies from the same movie we calculated the body threshold value for each movie individually. 

\begin{figure}
\begin{centering}
\includegraphics[width=.75\linewidth]{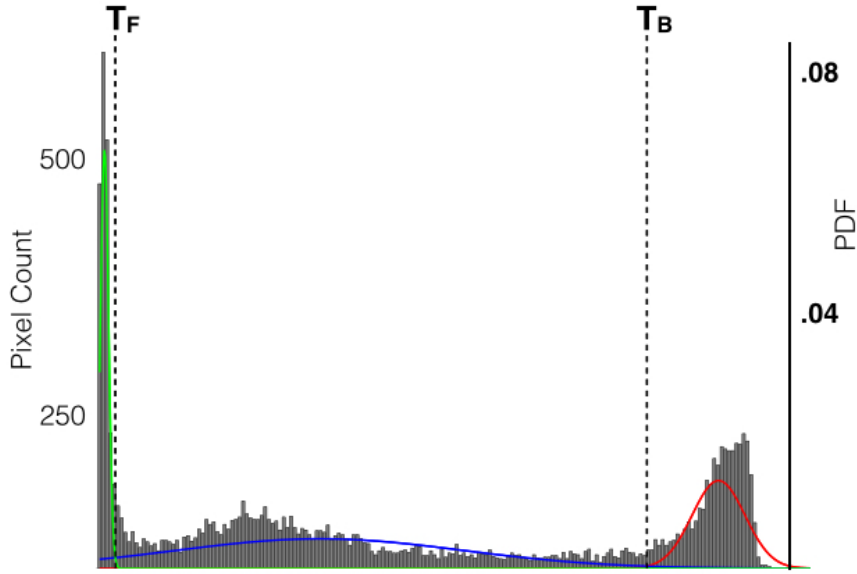}
\caption{ Histogram of pixel intensities within fly connected components for one sample image. The probabilities of belonging to each component of the gaussian mixture model (GMM) are shows in red (low intensity), blue (medium intensity), and green (high intensity). The thresholds for the image are determined by finding the intersection of the posterior probabilities between two adjacent sets of pixels given by the GMM and are labeled with dotted black lines.}
\label{Figure Hist}
\end{centering}
\end{figure}

Before segmentation and tracking begins, the entire movie is loaded into memory and 300 frames containing flies with completely separate silhouettes are sampled at random from the movie. Each of the sample images is loaded individually and a three component GMM (gaussian mixture model) is fit to all of the nonzero pixels that that make up the image (Fig. \ref{Figure Hist}). We used the Matlab function gmixdistribution.fit to estimate the means and mixing proportions of the three components using 10 replicates. At any given pixel value, we calculated the posterior probability of belonging to each of the three components. The posteriors were fit using a linear interpolation and the value at which a pixel is equally likely to be within the lowest- or middle-intensity component was found. This is the fly threshold $T_F$. The pixel value equally likely to be from the middle- or highest-intensity component is the body threshold $T_B$. We calculated $T_F$ and $T_B$ separately for each of  300  images sampled throughout the movie and the median threshold in each category is the final value used throughout the entire movie. 

\subsection{Assignment and Segmentation}\label{tracking}

\begin{figure}
\begin{centering}
\includegraphics[width=.75\linewidth]{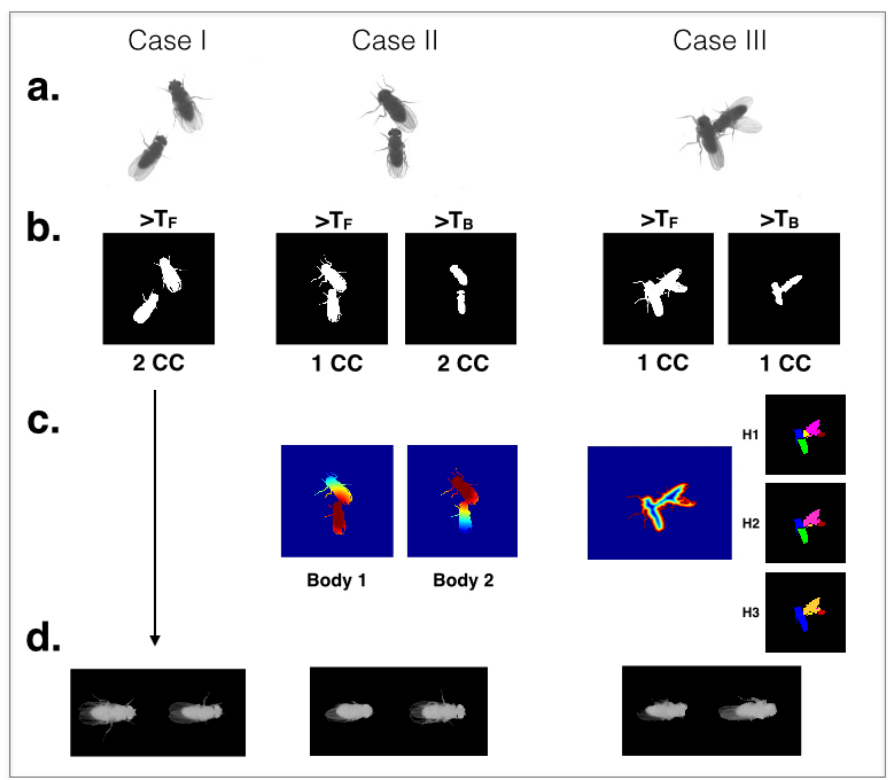}
\caption{Overview of algorithm used to create aligned movies of individual flies from original courtship movies. Examples of images created during particular steps in alignment and segmentation of a courting pair of flies depending on which heuristic was used. Case I: a) Original image, b) applying the lower threshold $T_F$ produces two separate connected components, d) output images are aligned masks of connected components from panel (b). Case II: a) Original image, b) applying threshold $T_F$ produces a single connected component, but isolated bodies of the two flies become apparent after applying threshold $T_B$, c) A gradient shows the distance of all post-threshold pixels to a given body mask. The pixels are assigned to whichever mask boundary they are closest to, d) output images are aligned masks of all pixels assigned to a given fly. Case III: a) Original image, b) thresholds $T_F$ and $T_B$ both produce images with a single connected component, c) the binary body image is produced by applying a watershed algorithm iteratively  with  increasing H-minima transform values until a maximum of four watershed regions are produced. The single largest region is assigned to a single fly mask while the other regions are grouped to produce the second fly mask, d) output images are aligned masks of all pixels assigned to that particular fly.}
\label{Figure 2}
\end{centering}
\end{figure}

To generate single-fly movies from paired-fly recordings, we need to segment the two flies from every image, assigning particular pixels to each fly, and track the motion of the flies over time. We broke this problem up into three cases. For each case, we used different segmentation methods based on how difficult flies are to track and identify in any given frame (Fig. \ref{Figure 2}). In the first case the flies are clearly separated in the arena. We segment the flies through a simple thresholding and object tracking procedure. In the second case the flies are touching, but their bodies are still separable by applying a simple threshold. In this case, identification based on size and centroid is still possible, but some method for assigning pixels between the individuals is necessary. The third case includes frames where fly appendages and bodies are touching and therefore requires heuristic methods to assign the pixels between the flies. This is the most rare case by far and usually occurs during copulation attempts.  Based on 15 movies of courting flies, corresponding to over 1 million frames, we find that Case I corresponds to $\sim90$\% of the frames, Case II to $\sim10$\% percent of the frames, and Case III to only $\sim0.25$\% of the frames. 

We begin tracking at the first frame in which flies are easily separated. Fly identity is assigned based on body area. For courting flies, we assume that the female is larger and assign identity based on the area of the body in the first frame. We analyze each frame in series and a threshold is applied to produce a binary image (\S \ref{threshold}). If there are two connected-components from this threshold, $T_F$, with areas greater than 500 pixels (a value easily larger than any accidental matter in the dish but smaller than a single fly) then each of these connected components is used as a mask for an individual fly (Case I). If there is only one connected-component from this threshold, but two components when using the larger "body'' threshold, $T_B$ (Case II), we assign wing and limb pixels to each fly based on distance to the body components. We then produce a mask for each fly using the assigned pixels. 

Finally, if there is only a single connected-component after thresholding with $T_B$ (Case III), then we must segment the body and assign pixels in another manner.  We first apply a binary distance transform to this connected component (using the bwdist command in MATLAB), which contains pixels from both the male and female flies. Pixel values in this image correspond to the distance to the closest edge of the connected component. In most cases, this image contains two large basins, corresponding to each individual fly, separated by a noisy ridge. Because copulation attempts normally happen with the male behind the female, this prior knowledge is incorporated when identifying these basins as the fly bodies. A watershed transform is then used to find these two basins \cite{meyer1994topographic}. In practice, however, the basins are noisy and contain a number of small basins that get segmented separately in the watershed. We solve this problem by iteratively applying an H-minima transform (imhmin in MATLAB) with an increasing cutoff depth until there are a maximum of four regions segmented by the watershed. Two watershed regions are rarely found in this step because the male's posture often includes spread wings which produces extra regions. For this reason we group all but the largest watershed region into a single mask and in this way create two individual regions. Pixels surrounding the combined watershed basins above the threshold $T_F$ are assigned to the different flies based on distance as in Case II. Since copulation is of great importance for the behavioral classification of courtship, this technique successfully tracks the flies even during highly occluded frames.

Assigning identity to each fly after the first frame depends not only on the size of each fly body but also on continuity from the previous frames. Each frame is considered both individually and in series so that prior centroid and identity information is always available. At times, segmentation distorts the body area of one or both flies, but assignment is maintained by assuming that the flies in a given frame will not travel far from their location in the previous frame and that fluctuations in body area are more likely than sudden changes to the fly position coordinates. The centroids and orientations of each fly in the original arena are recorded so that we can reconstruct behavior and analyze the effect of each measurement on the interaction behavior separately.

\subsection{Writing Aligned Movies}\label{movies}
We create a mean, or basis, image of each fly body and wing silhouette for use in alignment and scaling of the individual movies. These basis images are generated for each of the flies before tracking and segmentation begins. The bases are binary images that indicate the median size and shape of the fly body after thresholding and alignment of the unique individuals. Basis images are necessary because the paired flies are of different shapes and sizes and alignment is performed by matching a given fly image to an already aligned basis \cite{berman2014mapping}.  

As we segment each frame and assign fly identities, we also align the images. At each time point, we record the orientation angle of the fly, which is the value of angular rotation necessary to produce the best alignment. This is also used as a way to check the consistency of alignment. The previous orientation angle is used to check and correct fly alignment during segmentation that distorts or clips part of the fly body, when the normal algorithm may fail.

Tracking and segmentation values are recorded in increments of 1,000 frames. If the flies have become impossible to segment for a large number of frames, meaning that the third case of segmentation has been used for a large number of frames, tracking is terminated as it is assumed the flies have copulated. Each frame of the aligned movie for one of the interacting flies is 150x150 pixels and contains a centered and aligned fly. The area of the basis image for each fly body is calculated and used to scale each movie so that fly bodies for both flies consistently contain 1,500 pixels. This rescaling is necessary because all flies are slightly different in size and embedding onto the behavioral map will be more accurate when the final body shapes compared are as similar as possible. The two aligned and rescaled movies are saved separately as .avi files. 

\section{Results and Discussion}

\subsection{Behavioral Distributions for Isolated Flies and Courting Pairs}\label{maps}

\begin{figure}
\begin{centering}
\includegraphics[width=.75\linewidth]{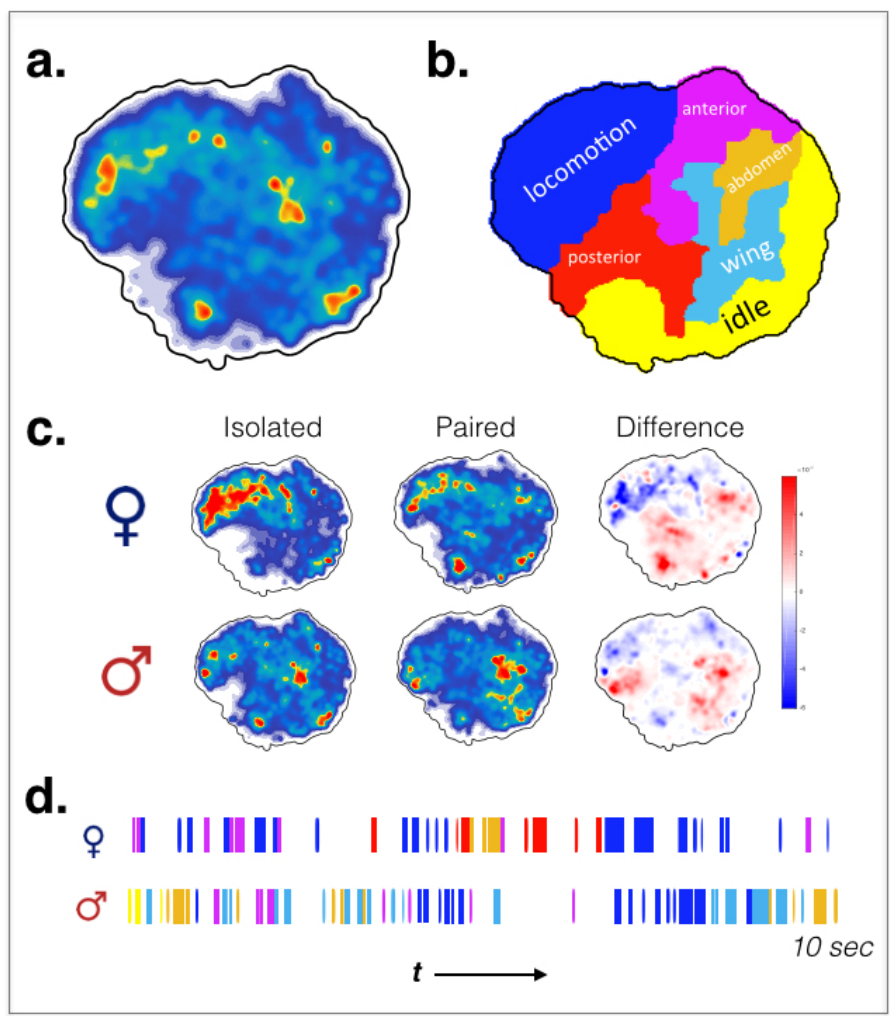}
\caption{Behavioral maps produced from Canton-S wild type flies in both isolated and courting contexts.  a) A density map containing all movies in the dataset (paired male, paired female, isolated male, isolated female).  b) Coarse descriptions of map regions based on visual inspection of the aligned movies.  c) Behavioral density maps created by plotting only points produced by courting males, isolated males, courting females, isolated females, and the differences between the isolated and paired contexts. Difference maps indicate which behaviors are enriched during courtship (red) or when individuals are isolated (blue). d) Simultaneous ethograms for a single ten-second bout of interaction with colors corresponding to the coarse labels in panel (b).}
\label{Figure 3}
\end{centering}
\end{figure}

To identify behaviors that are specific to courtship, we recorded separate movies of isolated males, isolated females, and paired males and females. Each aligned movie, whether from an isolated or paired individual, contains fly images of a standard size and can be put through our behavioral analysis pipeline to generate behavioral density maps. The final format of the aligned fly data is the same for each fly regardless of whether it came from an isolated or paired experiment. This allows us to combine and compare different sexes and conditions in order to create a common 2-dimensional space of all behavior seen across all experimental conditions. We use the pipeline described in Berman et al. to produce a single behavioral map that includes data from all isolated and paired flies. Briefly, this analysis includes dimensionality reduction of each image followed by a low-dimensional embedding of the temporal power spectrum \cite{berman2014mapping}  (Fig. \ref{Figure 3}). The behavioral map is a 2-dimensional histogram which provides a visualization of how much time flies spend performing any given behavior. The sharpness of certain peaks in the histogram indicates that some behaviors are more stereotyped than others. 

As described previously, the behavioral map naturally divides into large scale regions of similar behavior which we operationally label by visual inspection \cite{berman2014mapping, berman2016predictability}. These include locomotion, posterior behaviors such as rear grooming, anterior behaviors such antennal or eye grooming, actuation of the abdomen, and wing movements such as wing grooming and singing (Fig. \ref{Figure 3}b). These broad regions overlook much of the detail available in the map, but provide an entry point into making time-dependent comparisons of the behaviors of two interacting individuals. Ethograms, or time-dependent catalogues of the behaviors an individual performs, are produced by grouping points in the behavioral space as shown in Figure \ref{Figure 3}b and plotting the corresponding color assignment over time (Fig. \ref{Figure 3}d).  Parameters for setting the size of behavioral regions can be used to find reoccurring relationships and patterns by surveying the different properties of these outputs. This method can be used to find correlations between behaviors in communicating individuals over time, and to potentially locate behaviors that affect the outcome of courtship.

The embedding algorithm we use is stochastic, and produces a different map each time it is run, so we perform a single embedding using data from all individuals. A subsampling of reduced-dimensionality data from isolated males and females, as well as segmented males and females from the paired experiments was used to produce an initial behavioral embedding. Using many individuals guarantees that all present behaviors, regardless of how rare, will be represented in our behavioral map. By subsampling the space of behaviors so that all actions from any of the contexts are represented in our original embedding, we can re-embed additional data onto this map without having to calculate a new embedding from the beginning. As we are interested in comparisons across many different contexts, embedding all data into the same map allows for direct comparisons of the probability density, or the frequency with which flies perform particular actions, between subpopulations of the data  (Fig. \ref{Figure 3}c). 

We recorded the behavior of 12 isolated females, 12 isolated males, and 15 paired males and females. Comparing the behavioral distributions between contexts (isolated versus paired) reveals a number of features of courtship (Fig. \ref{Figure 3}c). Overall, females spend more time exhibiting fast locomotion while males produce many more wing-related motions regardless of the presence of other sex. Difference maps between the isolated and paired condition for each sex show how behavior is affected simply by the presence of the courtship partner. We find that females run less and display more wing movements when a male is present. On the other hand, males not only produce more wing extensions when the female is present, presumably indicative of courtship singing, but also show subtle shifts to other wing- and abdomen-related motions. We use the coarse behavioral descriptions here, but the same comparisons can be done at very specific points on the map as well, and this method may be used in the future to find the conditions in which subtle behaviors are enriched.

\subsection{Behavioral Dependence on Distance and Orientation}\label{position}
\begin{figure}
\begin{centering}
\includegraphics[width=.75\linewidth]{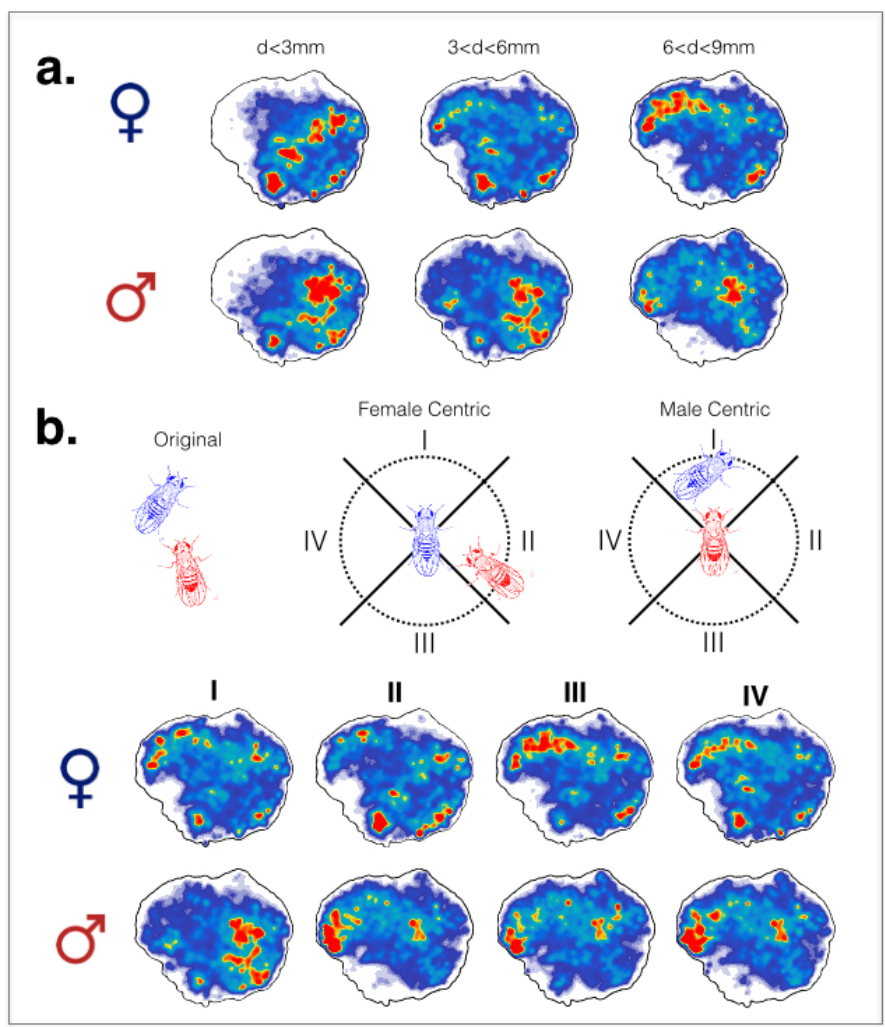}
\caption{Behavioral map densities for a variety of spatial contexts. a) Male and female behavioral density maps plotted for three intervals of the distance $d$ between paired flies ($<3$~mm (top), 3--6~mm (center), and 6--9~mm (bottom)). b) A schematic of distinct angular quadrants shows how flies may be viewed in the female- and male-centric spaces. Each row of behavioral maps corresponds to the behavioral density of the given sex while the other fly's angular position is in a given quadrant. }
\label{Figure 4}
\end{centering}
\end{figure}

We used the position of the tracked flies in space to calculate the following parameters: the distance from the centroid of the male fly to the centroid of the female fly, $d$,  the position angle of the female in a male-centered cartesian space, $\phi$, and the position angle of the male in a female-centered space, $\theta$ (Fig. \ref{Figure 1}b). These three values are enough to describe the relative orientation and distance of both flies, and make it simple to find the spatial context of a behavior from either individual at any point in time. We find that the behavior of both males and females varies as a function of the fly distance, ${d}$ (Fig. \ref{Figure 4}a). When ${d}$ is less than 3~mm, locomotion is underrepresented in both sexes. In contrast, wing, abdomen, and idle postures are much more likely. As distance increases ($3<d<6$~mm), overall locomotion at various speeds becomes more likely for the female while the male locomotes at a much higher relative speed. Due to the increase in locomotion, the likelihood of wing-, abdomen-, and idle motions diminishes. At larger separations ($6<d<9$~mm), The female spends almost all of her time locomoting whereas the male splits his time between fast locomotion and wing-based behaviors such as singing.

The behavior of each individual also depends on the relative orientation of their partner (Fig. \ref{Figure 4}b). We observe that the male wing extends, and therefore presumably sings, more when he is behind the female than in any other angular quadrant. This may be due to the mechanics of song propagation and detection by the female or simply the drive to keep up with the female when she is moving away from him. The female, on the other hand, is more likely to be in the the locomotion map region when the male is directly behind her. It is unknown whether these particular preferences are universal or specific only to this wild type strain of fly. By exploring these differences across strains and species, future work will determine whether these may be important selected characteristics of courtship.

\begin{figure}
\begin{centering}
\includegraphics[width=.75\linewidth]{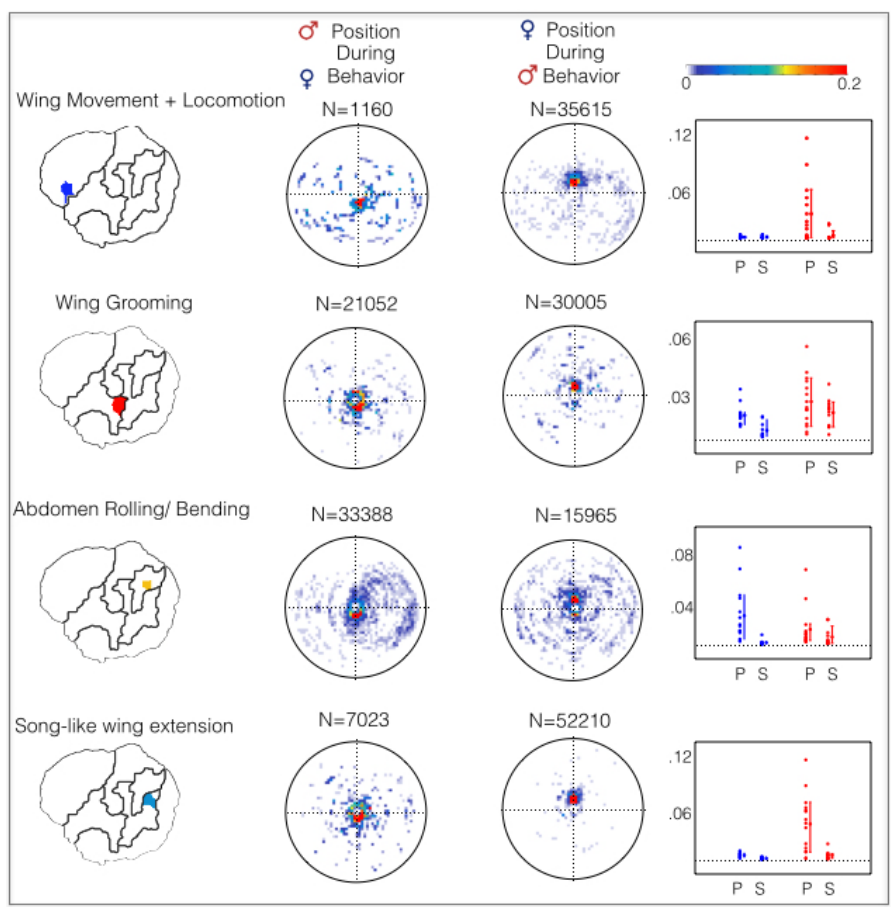}
\caption{A panel of sample behaviors displaying statistics on the location and orientation of the each fly when the behavior is performed relative to its partner, as well as the total fractional occupancy in the male and female during paired (P) and single trials (S). Each row describes the density of occupancy in space of the partnered fly while the specified individual performs a given behavior.  Behaviors specified are a) Fast locomotion with some wing involvement, b) rear and wing grooming, c) abdominal rolling or bending, and d) song-like wing extension.
}
\label{Figure 5}
\end{centering}
\end{figure}

\subsection{Fine Scale Dissection of Orientation Effects} 
We display fine behavioral resolution in another manner by creating probability density maps that describe the spatial organization of interacting flies during particular behaviors(Fig. \ref{Figure 5}). These plots reveal how the position of a partnered fly may selectively drive certain behaviors. The examples shown here are created by focusing on specific masked regions of the behavioral space, and then finding the spatial contexts in which those behaviors occurred. Each time an individual exhibits behavior in the masked region, the relative position of its partner is recorded. 

We find that locomotion with wing movements and song-like wing extensions (rows 1 and 4 of Fig. \ref{Figure 5}) are much more common in males, and especially in paired males. In fact, these behaviors are minimal in unpaired males and the frequency with which they are performed by females is low regardless of the presence or absence of a male. The same tight localization pattern in which the female is positioned directly in front of the male during male song-like wing extension indicates that male wing motions, and wing extensions in particular, are driven not only by gender and context of the behaving individual (male, and courtship), but also by the position of that individual's partner, the female. We also find that behaviors change in the paired female. Having a male present and courting induces an abdomen-bending behavior in the female (row 3 of figure Fig. \ref{Figure 5}). This action is also performed by the male, but its frequency is not dependent on the presence of a courtship partner. The relative spatial position of paired flies does not appear as crucial for abdomen related behaviors as it is for wing extensions. The paired females show a pronounced increase in abdomen bending over unpaired individuals, suggesting that this behavior is important for the communication of receptivity or rejection by the female in a courtship context. 

Finally, we see that even behaviors that are commonly displayed in both paired and unpaired conditions can still give insight about the mechanics of courtship. While wing grooming (row 2 of Fig. \ref{Figure 5}) occurs in all contexts for both sexes,  paired males and females both display an increase in this type of grooming over their unpaired counterparts. Further exploration of when grooming occurs, and in what spatial context, will aid in the understanding of how simple and common behaviors such as grooming may be employed to communicate during courtship. 

A compelling open question in behavioral science is one of how differences in complex traits arise from subtle changes in the genome. Equally compelling is the goal to understand how much of behavior is mediated by simple interactions with another individual, especially when these interactions are inevitable like in the case of courtship. We hope to use the method presented here to investigate these questions in fruit fly courtship by breaking down and quantifying this complex set of interactions. Here we have shown several ways to display behavioral differences between flies in different contexts. These methods are useful for interrogating courtship behavior in a principled matter, potentially leading to new insights on the interface of genes, environment, and behavior. 

\section{Experimental Methods}
\subsection{Fly Stocks and Experimental Conditions}
Wild type \textit{D. melanogaster} flies were isolated on eclosion and aged four to six days before imaging. Females were housed in batches of up to 50 virgin females while males were individually housed in 96 well deep well plates sealed with microporous tape. All flies were raised and imaged at 21-22C with a 12H light on/light off cycle. 

\subsection{Fly Behavioral Assays}
Our behavioral data is produced by filming silhouettes of a single or multiple flies from above in a circular chamber that is approximately 25mm across. Flies are prevented from crawling on the top of the chamber due to size restrictions and siliconizing reagent applied the day before filming. 

All behavioral experiments were filmed within 3 hours of the incubator light coming on. Single flies and courting fly pairs were filmed 12 at a time in three 4-camera setups after manual aspiration into individual arenas. Imaging was started approximately 5 minutes after introduction of flies, and flies were filmed continuously for 30 minutes. BIAS capture software was used to produce 1024x1024 pixel frames of each walking fly or courting fly pair at 100Hz.

\subsection{Preprocessing Movies and Coarse Fly Detection}\label{preprocessing}

In order to limit the size of files that are processed in series, and to begin the process of isolating touching flies, we first preprocess each movie by locating and saving the general region of the image where each fly is found in every frame. Each 1024x1024-pixel image from an input movie is subjected to coarse tracking to identify the objects. This process is completed in parallel by breaking up the long movie into short movies of 300 frames, where a threshold is applied to each frame and large connected-pixel components are discovered. If two separate connected objects are found, and their separation is larger than 100 pixels, a 150x150 region around each object is found and the regions are placed next to each other with blank space underneath to make up a 300x300-pixel image. If flies are separate but within 100 pixels of each other, a single 300x300-pixel image around the mean centroid of the flies is saved instead. If the image contains only one connected object then a single 300x300-pixel region around the centroid is saved. The short movies are recombined and the end result is a 300x300-pixel movie of the same length as the original movie comprised of either two single-fly images or a single double-fly image for each frame. 

The centroid coordinates of each recorded region are saved to an informational text file. This file is used later in describing the center of mass motion of each individual. This text file also includes information about which configuration the frame was saved in as a reference during segmentation. The preprocessed movie is saved as a .avi file and contains all of the information necessary, in conjunction withe the informational .mat file,  to reconstruct the original movie. This reduces the space requirements per movie, as well as creates a standard data type when using different acquisition software. 

\subsection{Copulation Detection}
Tracking and segmentation continues until the bodies have proved impossible to separate for over 500 frames. This cutoff doubles as a coarse copulation detector, and signals the end of tracking for that particular movie. It is possible to continue tracking once copulation has ended, but we do not consider these frames in our current comparisons as they are inherently different than pre-copulation movies when analyzing courtship behavior.  

\section{Acknowledgments}
This work was funded through awards from the National Institutes of Health (GM098090, GM071508), The National Science Foundation (IOS-1451197), the Howard Hughes Medical Institute through a Janelia Research Campus visitor project, and the Emory QuanTM Graduate Fellows Program.

\section{References}
\bibliographystyle {plain}
\bibliography {methodsPaper} 

\end{document}